\documentstyle[preprint,prb,aps]{revtex}
\renewcommand{\baselinestretch}{1.25}  
\begin{document}
\title{Nonlinear Transport through Coupled Double 
Quantum Dot Systems}
\author{R. KOTLYAR and  S. DAS SARMA}
\address{Department of Physics, University of Maryland, College Park,
Maryland  20742-4111}  
\date{\today}
\maketitle
\draft
\begin{abstract}
We investigate  sequential tunneling transport through a semiconductor 
 double quantum dot structure by combining a simple microscopic quantum 
confinement model with a Mott-Hubbard type correlation model. 
We calculate nonperturbatively the evolution of the Coulomb blockade 
oscillations as a function of the interdot barrier conductance, obtaining 
good qualitative agreement with the experimental data over the whole 
tunneling regime from the 
weak-coupling individual dot to the strong-coupling  
coherent double-dot molecular 
system.
\end{abstract} 
\newpage  

By tuning the tunnel barrier between the 
individual dots of a voltage-biased semiconductor double 
  quantum dot system, it has recently been possible\cite{waughprl}
 to 
observe the formation of an artificial double-dot molecule 
(with each dot as its atomic constituents) in Coulomb blockade 
transport experiments\cite{waughprl,crouch,livermore}. 
As the interdot tunneling is increased, the series of linear 
conductance peaks of the two 
individual near-identical dots\cite{waughprl,crouch,livermore} changes 
continuously to a series of 
split 
peaks which then form a well-defined Coulomb blockade 
oscillation series with twice the 
individual Coulomb blockade period.
This period doubling transition in the 
Coulomb blockade oscillations closely follows the energetics of the transition of two fully 
isolated dots into a single composite dot due to enhanced interdot 
tunneling. This transition 
raises important general questions on how the 
parameters which can be 
uniquely defined for the isolated system would renormalize in the 
transparent 
composite system. For example, to add an electron to a 
single dot requires energy in excess of the intradot 
interaction energy $u_{11}$. For two isolated dots in 
series in a 
double-dot system, supplying the required energy $u_{11}$ 
corresponds to the addition of one electron to each of the dots. 
With the increased `transparency' of the system due to enhanced 
tunneling, the energy
 $u_{11}$ 
required to add an electron changes as the doubling in the 
periodicity of the 
linear conductance Coulomb peaks demonstrates. The pertinent 
theoretical question is how to characterize the increase of the 
`transparency' of the composite system.
The classical capacitive charging model  
attributes this transition to the interdot electrostatic coupling energy $u_{12}$. 
In this model each dot is considered to contain an integer number of 
electrons. The splitting of the individual Coulomb blockade 
peak is proportional 
to the interdot coupling $u_{12}$. The saturation of the splitting in 
the strong
coupling limit is explained by the increase of $u_{12}$ to the 
value of $u_{11}/2$ with all the other 
system parameters assumed to be constant\cite{waughprl}.
The energy $u_{12}$ is taken to arise from the capacitive coupling 
between the 
two dots, which is classically determined by the fixed geometrical arrangement 
of the dots. Thus, within the classical capacitive charging model 
the fixed geometrical arrangement of the two dots in 
the experimental system provides no physical reason for the increase of 
$u_{12}$ necessary to account for the saturation of the observed 
peak splitting.  It has also been noted\cite{waughprl} that the 
classical capacitance values needed to 
quantitatively explain the observed peak splitting are unphysically high.

A recent fully quantum theory\cite{matveev,golden} of charge 
fluctuations between the two dots formulates the double-dot problem in 
terms of the dimensionless interdot conductance per tunneling channel 
$g=G_{int}/ (N_{ch}e^{2}/h)$ and the number of 
interdot tunneling 
channels $N_{ch}$. In the limit of the tunneling bandwidth being much 
larger than the 
intradot charging energy, each dot is treated as an infinite 
charge reservoir for the other dot. A perturbation analysis in 
$g$ or in $(1-g)$ reproduces the weak and strong interdot coupling 
limits of the peak splitting. This elegant perturbative analysis 
does not apply in the intermediate regime where 
$0 << g << 1$.  The charge fluctuation perturbation 
analysis has not yet been extended to the nonlinear transport regime, 
which is one focus of our work.

In this paper we study the transition from degenerate one 
dot Coulomb blockade oscillation to the coherent 
molecular double-dot oscillation by using a 2-site generalized 
Mott-Hubbard model within an extremely simple 
physically motivated microscopic confinement potential 
describing the double-dot system.
A Hubbard-type model of linear 
transport through the single-particle states of 
quantum dots\cite{stafford,chen,rkotlyarsfe} predicts the distinct 
phases in the conductance pattern characterized by an increase of the 
interdot tunneling strength $t$. In the strong tunneling limit the 
Mott-Hubbard 
insulator-metal transition opens a transmission channel through an array 
of quantum dots\cite{stafford}. This transition is equivalent to the 
formation of a single composite coherent molecular system 
with a concomitant doubling of Coulomb blockade periodicity.
To characterize the dependence of the Hubbard model
 parameters on the value of 
 the interdot conductance $g$,
we use a simple phenomenological step-well 
model for the confinement potential profile of the double-dot system. 
We obtain the exact 
Hubbard model predictions for the zero-bias voltage limit current peak 
splitting and the  threshold voltage for the 
onset of conduction through the 
double-dot with 1 and 2 excess electrons.

 The capacitive 
model of the experimental circuit configuration for the double-dot system \cite{waughprl,crouch,livermore} 
is shown in Fig. 1. We consider a symmetrical configuration of 
two identical GaAs dots with the same electrostatic couplings to the 
common back gate and to the bias leads, i.e. $C_{g1}= C_{g2}= C_{g}$, 
$C_{1}= C_{2}= C$, $V_{g1}=V_{g2}={V_{g}}$, and common 
self-capacitances $C_{0}$. The interdot capacitor 
with the capacitance $C_{int}$ provides the electrostatic 
coupling between the two dots. Following the experiments, we set an asymmetric 
bias across the system, i.e. $V_{1}=V, \ V_{2}=0$. 
We express the electrostatic part of the free energy of the
system using the classical capacitance matrix\cite{bookset} 
formalism. In the usual final 
step of going into the quantum-mechanical description we replace the 
classical excess charge on a dot by the charge density operator. 
Then the operator
of the electrostatic free energy of the coupled system of the double-dot 
and leads is 
\begin{equation}
\label{1}
F^{e}=u_{11} ({\hat{N}_{1}}^{2}+{\hat{N}_{2}}^{2})+
u_{12} {\hat{N}_{1}} {\hat{N}_{2}}+
eV(x_{b1} {\hat{N}_{1}}+ x_{b2} {\hat{N}_{2}})+
eV_{g}x_{g} ({\hat{N}_{1}}+ {\hat{N}_{2}}). 
\end{equation}
In Eq. (1) the Hubbard parameters are expressed through the elements of 
the capacitance matrix, i.e.
$u_{11}=e^{2} \frac{C_{\Sigma}}{2 \delta}, \ 
u_{12}=e^{2} \frac{C_{int}}{\delta}, \
x_{g}=\frac{Cg}{e^{2}}(2 u_{11}+u_{12}), \ 
x_{b1}=\frac{C}{e^{2}}(2 u_{11}), \
x_{b2}=\frac{C}{e^{2}}(u_{12})$,  
where 
$ C_{\Sigma}=C_{0}+C_{g}+C$, and 
$\delta= {C_{\Sigma}}^{2}- {C_{int}}^{2}$.
The bottom of the conduction band, considered to be the same in all 
GaAs electrodes in the system, is taken as the reference.  
Due to quantum confinement in the dot the continuous 
conduction band for 
an excess quasiparticle becomes a discrete series of 
 single-particle energy levels $\varepsilon_\alpha$ where $\alpha$ 
denotes the single-particle 
state including spin. We consider spin-degenerate 
single-particle levels in the dot. The quasiparticles are allowed to tunnel 
between the single-particle states in the two dots with 
 the tunneling amplitude $t_{\alpha}$.  
By including the kinetic energy in each dot and the tunneling 
energy, we write the total free energy 
operator as 
\begin{equation}
\label{2}
F_{0}=F^{e}+\sum_{i=1,2, \ \alpha} \varepsilon_{i\alpha} {{c}^{\dag}}_{i\alpha} c_{i\alpha}- 
\sum_{\alpha} (t_{\alpha} {{c}^{\dag}}_{1\alpha} c_{2\alpha} + h.c.). 
\end{equation}
The indices 1, 2 denote the spatial positions of dots, 
${\hat{N}}_{i}=\sum_{\alpha}{{c}^{\dag}}_{i\alpha} c_{i\alpha}$ is 
the density operator, where  ${{c}^{\dag}}_{i\alpha} (c_{i\alpha})$ 
is a creation (annihilation) operator for a quasiparticle on the i-th dot 
 in a state $\alpha$.

 The double-dot is isolated 
from the leads so that it is coupled to them only electrostatically and 
through (very weak) tunneling matrix elements $t^{1,2}_{\alpha}$.
 The conductance of the dot to the leads 
$G_{lead}=0.02 \  e^{2}/h \ (<< e^{2}/h)$ is kept constant throughout.
 (The lead-dot tunneling strengths $t^{1,2}_{1}$ 
are estimated to be $3 \ \mu eV$.)  
 The tunneling hybridization energy between the 
double-dot and the bias leads is treated as a weak 
 perturbation $H_{T}$: 
\begin{equation}
\label{3}
H_{T}^{1,2}=\sum_{k \alpha} (t^{1,2}_{\alpha} {{c}^{\dag}}_{k} c_{1,2\alpha} + h.c.),
\end{equation}
where the index $k$ denotes  a quasiparticle state in the leads.

Using Eqs. (1)-(3) it is possible to calculate the current through 
the double-dot system under the assumptions of sequential 
tunneling and weak coupling to the outside leads, provided that the level separation in the double-dot system is much less than the width of the transmission resonance. We skip the calculational details which follow 
the corresponding single dot Coulomb blockade 
formalism\cite{rkotlyarsfe,bookset,beenakker} (suitably generalized to the 
double-dot system). In particular, the Coulomb gap $\Delta V_{gap}$ 
and the normalized peak splitting\cite{waughprl}, $f$, defined as the ratio of the additional energy needed to increase the number of quasiparticles by one to its maximum (saturation) value are respectively given by (with the total number 
of particles being 1 or 2) 
\begin{equation}
\label{4}
\Delta V_{gap}=4 \left[
\frac{(2-b_{2})(u_{11}+\varepsilon_{1})-
\sqrt{b_{1}^{2}(u_{11}+\varepsilon_{1})^{2}+
[(2-b_{2})^{2}-b_{1}^{2}]t^{2}}}
{[(2-b_{2})^{2}-b_{1}^{2}]}\right],
\end{equation}
\begin{equation}
\label{5}
f=
 \left( u_{11}+u_{12}/2+2t-
\sqrt{(u_{11}-u_{12}/2)^{2}+2t^{2}} \right) / (U/2), 
\end{equation}
where $u$ and $t$ are those appearing in the Mott-Hubbard model defined 
through Eqs. (1) - (3),  $U$ is the intradot interaction energy of the 
isolated dot,  
$b_{1}= x_{b1} - x_{b2}$ and $b_{2}= x_{b1} + x_{b2}$. 
Thus a knowledge of the Mott-Hubbard parameters $u_{11}$, 
$u_{12}$, $t$ etc. allow us to obtain the complete current-voltage 
characteristics of the double-dot system. 

We construct a simple microscopic quantum mechanical model 
to phenomenologically describe the double-dot system depicted in Fig. 1. 
Our microscopic confinement model is shown as an inset of Fig. 2. 
The model uses two identical one dimensional infinite hard wall 
potential wells to describe the two isolated dots. Each dot is represented 
by a two-step well (as shown in Fig. 2) with the 
lower step well of width $a$ 
representing the intradot interaction energy of the individual dot and the 
rectangular barrier of potential height $V_{b}$ and width $d$ representing 
the (variable) tunnel barrier separating the two dots. When the barrier 
$V_{b}$ is large (e.g. $V_{b} \ \rightarrow \infty$) the tunnel conductance 
is vanishingly small and the two dot system is in the uncoupled `atomic'
limit whereas for small $V_{b}$ (e.g. $V_{b} \ \rightarrow 0$)
the dimensionless `tunnel' conductance approaches unity and the system 
is in the composite `molecular' limit. Within our one dimensional confinement model the barrier of height $V_{b}$ and width $d$ 
approximates the intradot constriction. Such simple one dimensional potential 
confinement models have earlier been used\cite{prange} to study 
quantum tunneling characteristics in three dimensional systems. 
Without any loss of generality we choose $d/a=0.1$ throughout our 
calculation -- the exact values of $d$ and $a$ are, of course, unknown 
but the only relevant point for our simple model to be 
qualitatively meaningful is for $d << a$. 

We evaluate the Mott-Hubbard 
interaction parameters $u_{11}$ and $u_{12}$ by taking the appropriate 
expectation values of the screened Coulomb interaction using the potential 
confinement model. (We take the screening length as 220A$^{o}$.)  
 The short-ranged part of the Coulomb interaction 
is assumed unscreened and is approximated by a delta function potential. 
This particular way of calculating the intra- and inter- dot Coulomb interaction energies is obviously 
{\it{not}} unique, and one can use alternative definitions 
for $u_{11}$ and $u_{12}$, which would give somewhat different 
numerical values. We feel, however, that within our simple confinement model our physically motivated definitions for the Mott-Hubbard parameters 
is reasonable. Similarly, the hopping parameter $t$ can be defined in several 
alternative ways within our microscopic model. We evaluate $t$ as 
$t=\Delta_{sas}/2$ where $\Delta_{sas}$ is the so-called 
symmetric-antisymmetric energy gap between the two lowest single particle energy levels in our model double-well potential. Conduction in 
the coupled dot system occurs through a single spin-degenerate quasiparticle state at the ground state energy $\varepsilon_{1}$ of the potential well. 
Finally, we need to evaluate the interdot conductance $g$ within our microscopic model in order to make direct contact between the experimental 
\cite{waughprl,crouch,livermore} (and earlier theoretical\cite{matveev,golden}) results and our model calculations. 
The interdot conductance $g$ can of course be exactly 
evaluated  for our 
simple one dimensional rectangular barrier model of the point contact separating the two dots. Because of our hard wall potential confinement 
model, however, the exactly calculated (`hard') conductance for a 
tunnel barrier of height $V_{b}$  and width $d$ is a rather poor 
approximation (even on a qualitative level) for the experimental interdot point contact 
conductance. We have, therefore, also employed a `soft' conductance 
approximation using a WKB expression, 
$g \approx \exp\{-2[\frac{2md^{2}}{\hbar^{2}}
(V_{b}-\varepsilon_{1})]^{1/2}\}$, which we believe better represents 
(on a qualitative level) the adiabatic confinement potential 
expected in the experimental double-dot system. We find much better qualitative agreement between our theory and experiment 
using the `soft' (as against the `hard') conductance model, which is what we will 
mostly present in this paper.

Within our highly simplified microsopic model for the double-dot 
system, the gate voltage induced lowering of the interdot barrier 
$V_{b}$ causes the crossover from two isolated dots (for 
large $V_{b}$) of size `$a$' each separated by a distance `$d$' to 
a single composite coherent  double-dot of size $2a+d$. Thus 
a single tunable parameter, $V_{b}$, controls the `transparency' 
of the system and causes the transition. Instead of using $V_{b}$ as the 
control parameter, however, we follow the experimental procedure 
of using the interdot tunnel conductance $g$ (determined completely by 
$V_{b}$ in the `soft' and `hard' approximations as described above) as the control parameter in depicting our results. As $V_{b}$ is 
tuned all the parameters of our model 
(e.g. $t, \ \varepsilon_{1}, \ u_{11}, \ u_{12}, \ g$) vary as known 
functions of $V_{b}$. We fix the individual dot size 
$a$ ($\approx 350 $ nm) using the experimental\cite{waughprl} value 
of the intradot interaction energy $u_{11} \approx \ 230$  $\mu$eV for 
the isolated dot. 
 
Our calculated Hubbard model parameters 
(e.g. $ \varepsilon_{1}, \ t, \ u_{11}, \ u_{12}$)
 are shown as functions of the corresponding interdot 
`soft' conductance $g$ in Fig. 2. Although $u_{11}$ and 
$u_{12}$ approach each other as $g$ increases, the simplicity and 
the inadequacy of our microscopic model does not produce 
$u_{11}=u_{12}$ for $g=1$. This is mainly due to the various 
approximations used in calculating the overlap integrals 
for the Coulomb energy. This discrepancy is the most severe 
quantitative limitation of our model. 

In Fig. 3 we show our calculated Coulomb Blockade oscillations 
for the double-dot system (in the linear regime) for four values of the 
`soft' interdot conductance $g= .16, \ .52,\  .8$, and
$.99$ at $T=87$ mK ($k_{b}T << u_{11}$). Note that for the sequential tunneling situation considered here, the width of the Coulomb blockade peaks arises entirely from thermal broadening. Our calculated evolution of the Coulomb blockade oscillations from the degenerate single-dot oscillations (at low $g$) through peak splitting (intermediate $g$) to the 
eventual period doubling (at large $g$) of the Coulomb blockade oscillations is qualitatively similar to experimental observations (cf. Fig. 5 of Ref. 1). 
To further quantify our results we show in Fig. 4 our calculated 
normalized 
peak splitting $f$ (Eq. 5) and the Coulomb gap $\Delta V_{gap}$ (Eq. 4) 
as functions of the interdot tunnel conductance $g$, both for 
`soft' and `hard' conductance models. The corresponding experimental
data\cite{waughprl} for $f$ show considerable scatter and our results 
(for the soft model) agree with experiment. The `hard' model, however, 
disagrees with the experimental results for reasons discussed above. We 
 point out (and this has been alluded to above) that the main 
quantitative limitation of our model seems to be a weaker (stronger) dependence of both $f$ and $\Delta V_{gap}$ on the tunnel conductance 
$g$ for small (large) values of $g$ than seen in experiments. 

Finally, in Fig. 5 we show for $g$ = .16, . 8 and .99 our calculated nonlinear 
Coulomb blockade transport characteristics for the double-dot 
system by plotting the calculated current (in gray scales) as a function 
of both the source-drain voltage and the gate voltage. Again, our results 
are in good qualitative agreement with the experimental data\cite{crouch} 
with the main quantitative discrepancy arising from the inadequacy 
in our tunnel conductance value $g$ which is higher than the corresponding 
experimental result. 

In conclusion, using a simple 
single-parameter ($V_{b}$) 
one dimensional microscopic confinement 
model we calculate 
{\it{nonperturbatively}} the linear and nonlinear Coulomb blockade 
characteristics of a double-dot system 
as a function of the interdot tunnel conductance. 
 Our results are in reasonable qualitative agreement 
with the experimental results which is really the best we 
can hope for given the great simplicity of our model. 
We can obtain  
better quantitative agreement with experiment by using 
additional (e.g. $d$ and $a$) adjustable parameters,
but we feel that to be not particularly 
meaningful within our simple model.

This work was supported by the US-ONR.
The authors thank C. H. Crouch for helpful discussions of her two-dot 
transport experiments.

\renewcommand{\baselinestretch}{1.05}
\begin{figure}
\caption{
The equivalent circuit of the double-dot system under study. 
 The values of the capacitances 
$C_{1}=C_{2}=38 \ aF$. The other parameters are defined 
in the text. 
}
\end{figure}
\begin{figure}
\caption{
Variation of the Hubbard model parameters
$u_{11}, \ u_{12}, \ t,$ and $\varepsilon_{1}$ with
`soft' interdot conductance $g$. The conductance 
$g$ is in units of $2e^{2}/h$. 
Inset: The
 step-well model 
as defined in the text.
}
\end{figure}
\begin{figure}
\caption{
The conductance $g_{dd}$ in units of $e^{2}/h$ 
through the double quantum dot system 
 versus the gate voltage $V_{g}$ 
for four values of the
`soft' interdot conductance $g= .16, \ .52,\  .8$, and
$.99$ at $T=87$ mK in the linear regime ($V_{sd}=10$  $\mu$eV). 
The full thermal equilibration in the double dot system 
is assumed before and after each  
tunneling event. We use
$\varepsilon_{2}/ \varepsilon_{1}= 2, \  t_{2}/t_{1}=1.2$ and
$t^{1,2}_{2}/t^{1,2}_{1}=1.2$ for the single-particle spin degenerate
levels.}
\end{figure}
\begin{figure}
\caption{
Calculated
normalized
peak splitting $f$ (Eq. 5) and the Coulomb gap $\Delta V_{gap}$ (Eq. 4)
as functions of the interdot tunnel conductance $g$
(top and bottom panels) for `soft' and `hard' (thick 
lines) conductance models. The dashed lines show the boundaries 
for the scatter of the experimental data (taken from  Fig. 5 of 
the second paper in Ref.1). 
The top thin line in the bottom panel shows 
the Coulomb gap for {\it{fixed}} values of $\Delta V_{gap}$ with   
$u_{11}=227  \ \mu eV$ and $\ u_{12}=.11 \ \mu eV$.
}
\end{figure}
\begin{figure}
\caption{
The nonlinear current $I$ through the double-dot system 
(at $T=87 \ mK$)
 is plotted 
versus
the values of the gate voltage $V_{g}$
and the bias voltage $V_{SD}$ for three
values of the interdot `soft' 
conductance  $g = 0.16, \ 0.8, \ 0.99$.
The brightest shades in the plots  correspond to 
$I=59, \ 76, \ 78 \ pA$ 
 for the graphs from top to bottom. The states with $N=0$ to $N=8$ electrons 
in the double dot system contribute to $I$ in the shown parameter space. 
}
\end{figure}
\end{document}